\documentclass[conference, a4paper]{IEEEtran}
\IEEEoverridecommandlockouts
\usepackage{cite}
\usepackage{amsmath,amssymb,amsfonts}
\usepackage{algorithmic}
\usepackage{graphicx}
\usepackage{textcomp}
\usepackage{xcolor}
\usepackage{url}
\usepackage{soul}
\usepackage{booktabs}
\usepackage{tabularx}
\usepackage{dblfloatfix}

\def\footnoterule{\relax%
  \kern-5pt
  \hbox to\columnwidth{\vrule width\columnwidth height 0.2pt}
  \kern5pt}
\makeatother

\begin{document}

\title{OpenSky Report 2025: Improving Crowdsourced Flight Trajectories with ADS-C Data}
\newcommand{\specialcell}[2][c]{%
  \begin{small}%
  \begin{tabular}[#1]{@{}c@{}}%
 #2%
  \end{tabular}%
  \end{small}}

\author{%
Junzi Sun\IEEEauthorrefmark{1}\IEEEauthorrefmark{2}, %
Xavier Olive\IEEEauthorrefmark{1}\IEEEauthorrefmark{3}, %
Martin Strohmeier\IEEEauthorrefmark{1}, %
Vincent Lenders\IEEEauthorrefmark{1}%
\\[1em]
\begin{tabular}{ccc}%
\specialcell{\IEEEauthorrefmark{1}\small OpenSky Network\\Burgdorf, Switzerland} &
  \specialcell{\IEEEauthorrefmark{2}\small Faculty of Aerospace Engineering \\ Delft University of Technology\\ Delft, the Netherlands}
  \specialcell{\IEEEauthorrefmark{3}\small ONERA -- DTIS\\Universit\'e de Toulouse\\Toulouse, France} &
   
\end{tabular}
}

\maketitle

\begin{abstract}
The OpenSky Network has been collecting and providing crowdsourced air traffic surveillance data since 2013. The network has primarily focused on Automatic Dependent Surveillance--Broadcast (ADS-B) data, which provides high-frequency position updates over terrestrial areas. However, the ADS-B signals are limited over oceans and remote regions, where ground-based receivers are scarce. To address these coverage gaps, the OpenSky Network has begun incorporating data from the Automatic Dependent Surveillance--Contract (ADS-C) system, which uses satellite communication to track aircraft positions over oceanic regions and remote areas. In this paper, we analyze a dataset of over 720,000 ADS-C messages collected in 2024 from around 2,600 unique aircraft via the Alphasat satellite, covering Europe, Africa, and parts of the Atlantic Ocean. We present our approach to combining ADS-B and ADS-C data to construct detailed long-haul flight paths, particularly for transatlantic and African routes. Our findings demonstrate that this integration significantly improves trajectory reconstruction accuracy, allowing for better fuel consumption and emissions estimates. We illustrate how combined data captures flight patterns across previously underrepresented regions across Africa. Despite coverage limitations, this work marks an important advancement in providing open access to global flight trajectory data, enabling new research opportunities in air traffic management, environmental impact assessment, and aviation safety.
\end{abstract}


\section{Introduction}\label{sec:intro}

Since the increasing adoption of the Automatic Dependent Surveillance--Broadcast (ADS-B) system from the early 2000s, the aviation industry has experienced a significant increase in flight tracking data availability. ADS-B has become the standard for tracking aircraft positions, providing real-time updates on flight paths, speeds, and altitudes. The OpenSky Network, established in 2013 as a global crowdsourced sensor system, has played a crucial role in democratizing access to this data for researchers worldwide. Over the past decade, OpenSky has published six comprehensive reports covering diverse topics, including traffic collision avoidance systems \cite{schafer2019opensky}, emergency situation analysis \cite{olive2020opensky}, fleet renewal \cite{sun2021opensky}, sustainability studies \cite{sun2022opensky, sun2024opensky}, and additional data sources such as FLARM \cite{olive2023opensky}. This open data initiative has motivated extensive research in air traffic management, environmental impact assessment, and aviation safety, resulting in nearly 600 academic publications.

Despite these advances, a significant challenge persists in constructing complete flight trajectories: ADS-B coverage is inherently limited to regions with sufficient ground-based receivers. This limitation creates substantial data gaps over oceanic regions, the global South, and sparsely populated areas. These gaps are particularly problematic for research requiring continuous trajectory data, such as fuel consumption analysis, emissions modeling, and optimal route planning. Previous approaches often relied on simplistic great circle approximations between ADS-B coverage zones, introducing potentially significant errors in trajectory-dependent analyses.

An alternative approach to address these coverage gaps is space-based ADS-B, which uses satellite constellations to detect aircraft ADS-B signals globally, including over oceans and remote regions. Companies like Aireon and Spire have deployed receivers on the Iridium NEXT satellite constellation, offering near-global surveillance capabilities. In previous study, through the collaboration with Spire, we have demonstrated the potential of space-based ADS-B data for aviation research \cite{sun20232evaluating}. However, space-based ADS-B comes with some limitations for research applications. The data is proprietary and typically available only through commercial licensing agreements. These agreements often include restrictive terms that prevent open sharing and republication of the data, limiting collaborative research and independent verification of results. While space-based ADS-B represents a technological solution to coverage limitations, its commercial nature creates barriers to the open science principles that have driven innovation in the aviation research community.

To address these current limitations, the OpenSky Network has begun incorporating data from the Automatic Dependent Surveillance--Contract (ADS-C) system. Unlike ADS-B, which relies on ground-based receivers, ADS-C leverages satellite communication to provide position updates over oceanic and remote regions where ADS-B signals are unavailable. In an earlier exploratory study, we demonstrated the potential of ADS-C data by reconstructing a limited set of transoceanic trajectories \cite{xapelli2023first}, providing the first openly available data of this kind for aviation researchers.

The difference between ADS-B and ADS-C lies in their communication infrastructure. While ADS-B broadcasts position data to any receiver within range, ADS-C establishes a contract between the aircraft and specific air traffic service units (ATSUs), transmitting data via satellite networks such as Inmarsat. This communication method enables ADS-C to operate beyond line-of-sight limitations, though at lower update frequencies (typically every 10 to 15 minutes, compared to 0.5 to 2 seconds for ADS-B). The C-Band Inmarsat satellites used for ADS-C transmission are geostationary and collectively provide extensive global coverage. In this study, we focus specifically on data from the Alphasat satellite, which covers Europe, Africa, and substantial portions of the Atlantic and Pacific Oceans.

This paper presents a comprehensive approach to combining the complementary strengths of ADS-B and ADS-C data. Our methodology integrates high-frequency ADS-B position updates from terrestrial areas with ADS-C updates over oceanic and remote regions. Using nearly a year's worth of ADS-C data collected in 2024 (comprising over 720,000 messages from 2,606 unique aircraft), we develop data fusion techniques that overcome the challenges of differing update rates and resolution between the two systems. Our approach connects ADS-B coverage boundaries with ADS-C oceanic segments, generating continuous trajectories that provide unprecedented insight into flight paths across previously data-sparse regions.

The combined dataset enables significant advancements in several research areas. It allows for more accurate fuel consumption and emissions calculations by replacing estimated great circle routes with actual flight paths. It provides new insights into flight patterns across the African continent, where traditional ADS-B coverage has been limited. Furthermore, it creates opportunities for analyzing the impact of upper wind patterns on transatlantic routing decisions and optimizing flight paths for reduced environmental impact.

The remainder of this paper is structured as follows: Section \ref{sec:sources} discusses the sources of ADS-B and ADS-C data, while Section \ref{sec:characterization} provides a detailed characterization of the ADS-C dataset. Section \ref{sec:data_fusion} describes our methodology for combining two different data streams. Section \ref{sec:applications} presents applications of the enhanced dataset, focusing on emissions quantification and African flight analysis. Finally, Sections \ref{sec:limitations} and \ref{sec:conclusion} discuss limitations and conclude with directions for future work.

\section{Flight Trajectory Data Sources}\label{sec:sources}

\subsection{The OpenSky Network}

The OpenSky Network is a crowdsourced sensor system that collects surveillance data for air traffic control (ATC) applications. Its main goal is to offer the public access to vast, unfiltered ATC data while also supporting the advancement of ATC technologies and procedures through research. Since 2013, the network has steadily gathered air traffic surveillance data. Unlike commercial flight tracking services such as Flightradar24 or FlightAware, the OpenSky Network preserves the original Mode~S replies received by its sensors in a comprehensive historical database that researchers and analysts across multiple disciplines can utilize.

Initially, the network operated with eight sensors located in Switzerland and Germany. Today, it has expanded to over 7,000 registered receivers spread across the globe. As of 2025, OpenSky's dataset contains more than a decade's worth of ATC communication data. Although the network began by focusing exclusively on ADS-B, it extended its data capture to include the Mode~S downlink channel in March 2017. Recently, it also added other technologies such as FLARM \cite{olive2023opensky} and VHF. The current dataset now includes more than 40 trillion Mode S replies.

Over the years, the OpenSky Network has seen consistent growth and development, noting the introduction of dump1090 and Radarcape feeding solutions as well as the use of non-registered, anonymous receivers. This practice, however, was halted in early 2019 to ensure a consistent quality of feeder data. In March 2020, there was an approximate 30\% drop in daily flights compared to previous levels, mirroring the global reduction in air travel due to the COVID-19 pandemic. The processing of messages by the OpenSky Network has been improved to eliminate duplicate messages received by multiple sensors, thereby avoiding a misinterpretation that traffic levels never returned to pre-pandemic values.

The network's global data collection depends entirely on its crowdsourced receivers, in which enthusiasts, academics, and various supporting institutions primarily operate. The range of each sensor is confined by the line-of-sight of its antenna, typically around 400–500 km for top-performing antennas, reaching the radio horizon. The regions where these networks grow most organically generally reflect areas that are densely populated and wealthier. Between 2018 and 2025, the network's global coverage reached a saturation point common to many crowdsourced systems, with most new sensors improving reception only at lower altitudes in regions already covered in Europe, the US, and other developed areas. Nonetheless, significant coverage improvements can still be seen in the Middle East, South Asia, and New Zealand.

In addition to the payload of each Mode~S downlink transmission, OpenSky stores extra metadata, including precise timestamps (which are helpful for multilateration), receiver location, and signal strength, which varies with the receiver hardware. Further details regarding OpenSky's history, architecture, and various use cases can be found in our previous reports \cite{sun2024opensky,olive2023opensky,sun2022opensky,schafer2019opensky,olive2020opensky,sun2021opensky,strohmeier2017large}.\footnote{The website and data are available under  \url{https://opensky-network.org}}

\begin{table*}[!b]
\centering
\caption{ADS-C Report tags and their parameters}
\label{tab:adsc_tags}
\begin{tabularx}{\linewidth}{cll}
\toprule
\textbf{Tag No.} & \textbf{Report Type} & \textbf{Parameters} \\
\midrule
03 & Acknowledgement & Contract Number (e.g., 1, 2, 3) \\
\midrule
04 & Negative Acknowledgement & Contract Request Number, Reason \\
\midrule
05 & Noncompliance Notification & Contract Number \\
\midrule
07 & Basic Report & Latitude, Longitude, Pressure Altitude, Time of Position Report \\
\midrule
09 & Emergency Basic Report & Latitude, Longitude, Pressure Altitude, Time of Position Report, Emergency Status \\
\midrule
10 & Lateral Deviation Change Event & Latitude, Longitude, Pressure Altitude, Time of Event, Lateral Deviation \\
\midrule
12 & Flight ID & Flight callsign \\
\midrule
13 & Predicted Route & Next Waypoint, ETA and Altitude + Next+1 Waypoint, ETA, and Altitude \\
\midrule
14 & Earth Reference Data & True Track, Ground Speed, Vertical Speed \\
\midrule
15 & Air Reference Data & True Heading, Mach Speed, Vertical Speed \\
\midrule
16 & Meteorological Data & Wind Speed, True Wind Direction, Temperature \\
\midrule
17 & Airframe Identification Group & 24-bit ICAO transponder code \\
\midrule
18 & Vertical Rate Change Event & Latitude, Longitude, Pressure Altitude, Time of Event, Vertical Rate Change \\
\midrule
19 & Altitude Range Event & Latitude, Longitude, Pressure Altitude, Time of Event, Altitude Range Deviation \\
\midrule
20 & Waypoint Change Event & Latitude, Longitude, Pressure Altitude, Time of Event, New Active Waypoint \\
\midrule
22 & Intermediate Projected Intent Group & Distance, True Track, Altitude, ETA \\
\midrule
23 & Fixed Projected Intent Group & Latitude, Longitude, Altitude, ETA \\
\bottomrule
\end{tabularx}
\end{table*}

\subsection{Automatic Dependent Surveillance - Contract (ADS-C)}

\begin{figure}[htb]
    \centering
    \includegraphics[width=1\linewidth]{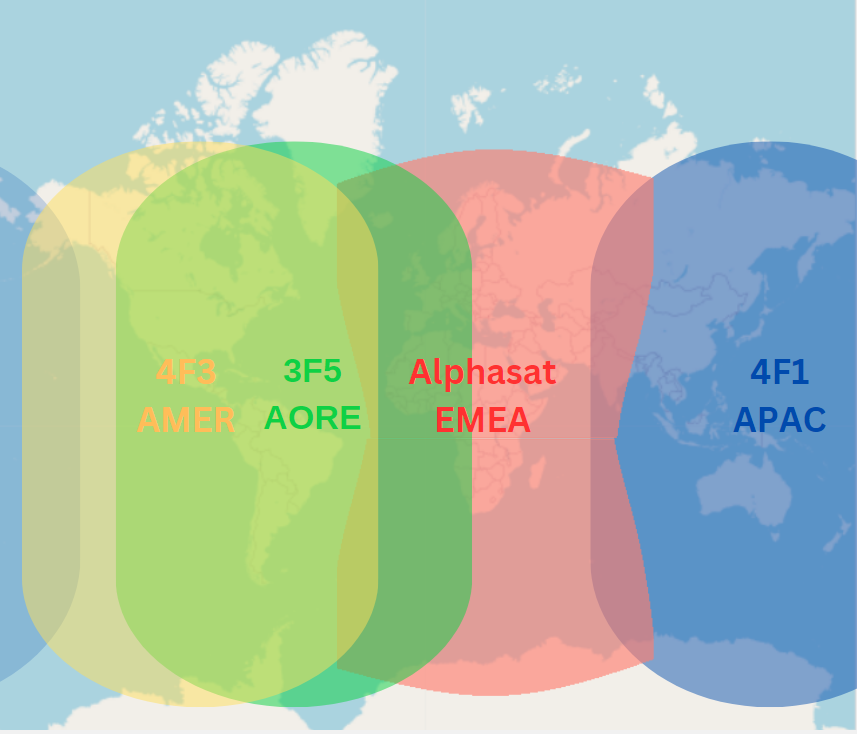}
    \caption{Distribution of ADS-C satellite coverage globally through the Inmarsat satellites \cite{xapelli2023first}}
    \label{fig:inmarsat}
\end{figure}

ADS-C is now an appealing source of data for several online flight trackers such as Flightradar24, ADSBExchange, or Airplanes.live. In 2023, OpenSky has also begun incorporating ADS-C into its offerings \cite{xapelli2023first}. At the core of the protocol stack lies the ADS-C message. As previously mentioned, ADS-C stands for Automatic Dependent Surveillance--Contract and, as such, is very similar to ADS-B in meaning and principal utility for ATC. In contrast to being \textit{Broadcast} by every plane's transponder to anyone in the radio range, ADS-C connects directly to ATC only and does so via several satellite transport layers (Inmarsat and Iridium). In this work, we only deal with ADS-C over Inmarsat, in particular, the Alphasat satellite covering Europe, Africa, and large parts of the Atlantic and Pacific Ocean (see Fig. \ref{fig:inmarsat}).

\textit{Contracts} refer to the agreements established between aircraft and ATSUs (Air Traffic Services Units) to exchange data. While an aircraft can establish contracts with multiple ATSUs simultaneously, messages are exchanged only between the aircraft and the ATSU with which it has a specific contract. This concept differs from ADS-B, where aircraft broadcast their messages to all receivers within range.

\textit{Contracts} are central to ADS-C, as they govern the transmission of surveillance data from aircraft to ATSUs. To initiate a contract, the ATSU sends a request specifying the required surveillance data. The type of contract established determines the data relayed by the aircraft:

\begin{itemize}
    \item \textbf{Periodic Contract}: The ATSU requests ADS-C reports at set intervals. These reports include key information such as flight ID, predicted route, earth reference, meteorological data, airframe ID, air reference, and aircraft intent.
    \item \textbf{Event Contract}: The aircraft transmits reports when specific events occur, such as altitude changes or reaching a designated waypoint. Events can include vertical range changes, altitude range alterations, waypoint shifts, and lateral deviations.
    \item \textbf{Demand Contract}: The aircraft sends a single report upon request. This type of report is beneficial if a scheduled periodic report fails to arrive.
\end{itemize}

ADS-C contracts typically operate in \textit{normal mode}. However, they can also function in \textit{emergency mode}, which can be triggered either by the aircraft (through an emergency report) or by the ATSU (via an emergency contract). When emergency mode is activated, the aircraft transmits reports more frequently than in normal mode.

Each ADS-C message includes at least a basic report, which contains the aircraft's latitude, longitude, altitude, timestamp, and a figure of merit. The figure of merit indicates the accuracy of the positional data and whether the Traffic Alert and Collision Avoidance System (TCAS) is operational. Beyond this, ADS-C messages can include optional sub-messages or tags. There are 18 different tags used in the downlink format. The basic report is associated with tag 07. Additionally, tags 09, 10, 18, 19, and 20 provide a more detailed position report, while tags 13, 14, 15, 22, and 23 help predict the aircraft's future position. Table \ref{tab:adsc_tags} provides an overview of the different ADS-C report tags and their associated parameters.

It is also worth noting that ADS-C over Inmarsat represents just one of several methods for transmitting ADS-C data. Alternative channels include ACARS, VHF, HF, and Iridium satellite services. For this study, we analyze the ADS-C data collected through the Inmarsat satellite network. For a detailed discussion of ADS-C message structure and functionality, please refer to the relevant specifications in the documents ARINC~745~\cite{arinc-745} and DO-258A~\cite{do-258a}.

\section{Characterization of the ADS-C Dataset} 
\label{sec:characterization}

We used version 1.4 of the OpenSky ADS-C data \cite{xapelli_2025_14659997} as a basis for this analysis. It consists exclusively of messages that are sent to the Alphasat satellite. In the following, we characterize this dataset of 720,415 messages. In the entire dataset, there are around 2,600 unique aircraft based on the 24-bit ICAO transponder code. Compared to the total number of aircraft we see in this region from historical data (around 52,800), the ADS-C dataset represents a small fraction  (5\,\%) of the total number of aircraft. The statistics are shown in Table \ref{tab:dataset}.

\begin{table}[!htbp]
\centering
\caption{Characteristics of the ADS-C dataset}
\label{tab:dataset}
\begin{tabular}{lc}
\toprule
\textbf{Total Messages} & 720,415 \\
\midrule
\textbf{All Aircraft} & 52,800 \\
\midrule
\textbf{All ADS-C Aircraft} & 2,600 (5\,\%)\\
\bottomrule
\end{tabular}
\end{table}

\begin{figure}[!htbp]
    \centering
    \includegraphics[width=0.95\linewidth]{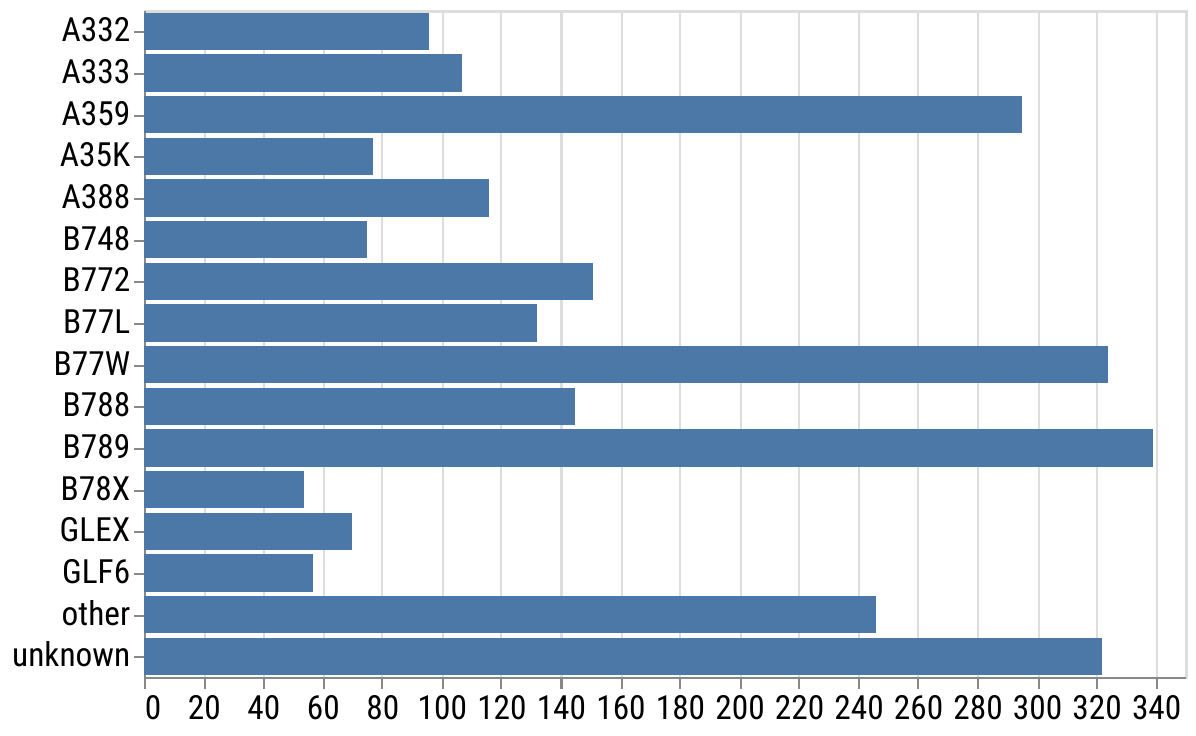}
    \caption{Distribution of the transponder type codes in the ADS-C dataset}
    \label{fig:transponder_typecode}
\end{figure}

Figure \ref{fig:transponder_typecode} illustrates the distribution of transponder type codes in the dataset. The vast majority of aircraft are long-haul commercial jets such as the Boeing 777, Boeing 787, Airbus A330, Airbus A350, and Airbus A380. A few business jets, particularly Gulfstream models, are also present in the dataset.

\subsection{Observed Message Tags}
\begin{figure}[!htbp]
    \centering
    \includegraphics[width=1\linewidth]{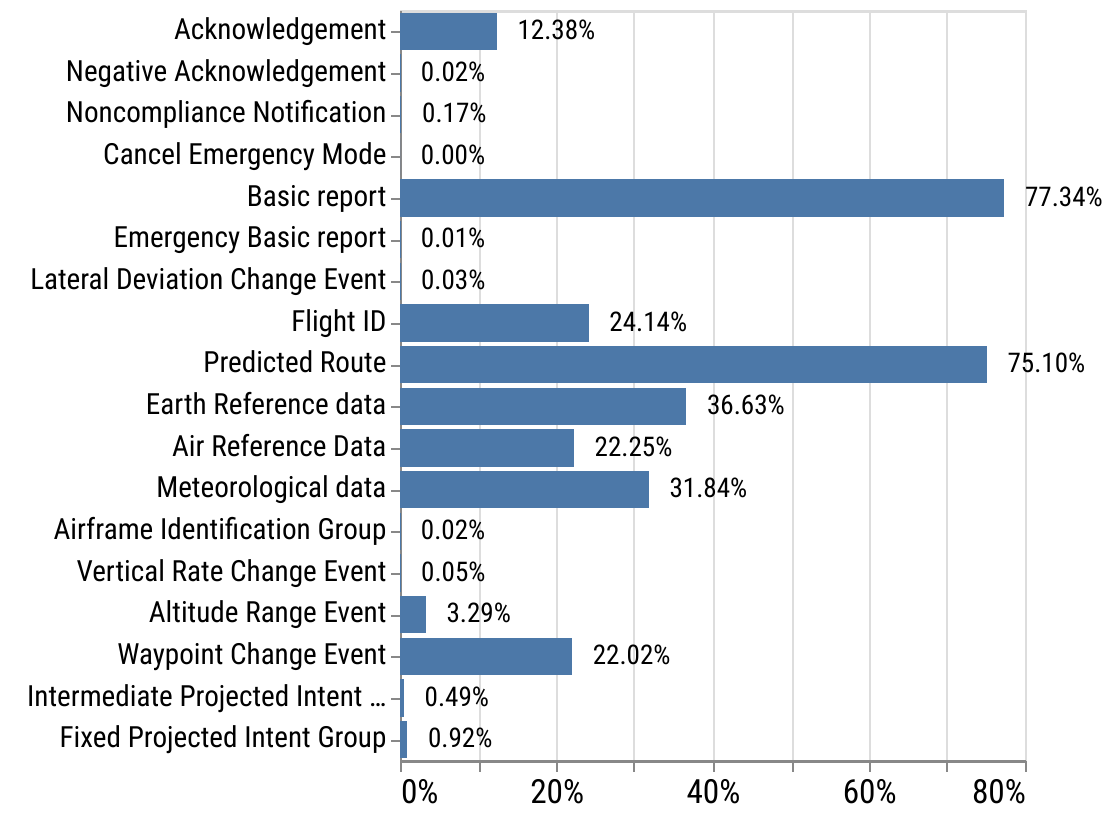}
    \caption{Distribution of ADS-C message tags in the  dataset. Multiple tags in a single message are possible, hence the numbers not summing to 100\%}
    \label{fig:message-stats}
\end{figure}

Figure \ref{fig:message-stats} shows the different message types received. As should be expected, basic reports are the most common. They are provided in almost 80\% of the data. Basic reports are closely followed by predicted route messages, also available in around 75\% of all received ADS-C reports. Five report types are reasonably common, available from about 20\% almost 40\% of the time, in descending order: Earth reference data, meteorological data,  air reference data, Flight IDs, and waypoint change events. The remaining event and message types are rare.

\subsection{Observed ADS-C Ground Stations}

\begin{figure}[!htbp]
    \centering
    \includegraphics[width=1\linewidth]{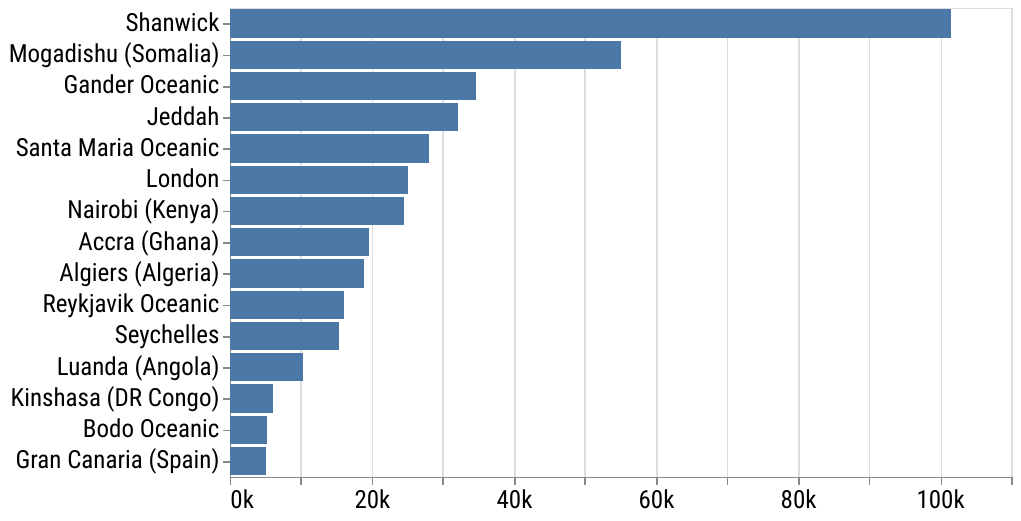}
    \caption{Most popular ATSUs.}
    \label{fig:atsu}
\end{figure}

The ADS-C messages in the dataset are routed via different ATSUs within the satellite range. Figure \ref{fig:atsu} shows the most used ATSUs in descending order. Shanwick leads with over 100,000 messages, followed by Mogadishu and Gander Oceanic. 

\begin{figure}[!htbp]
    \centering
    \includegraphics[trim=80 0 80 10, clip,width=\linewidth]{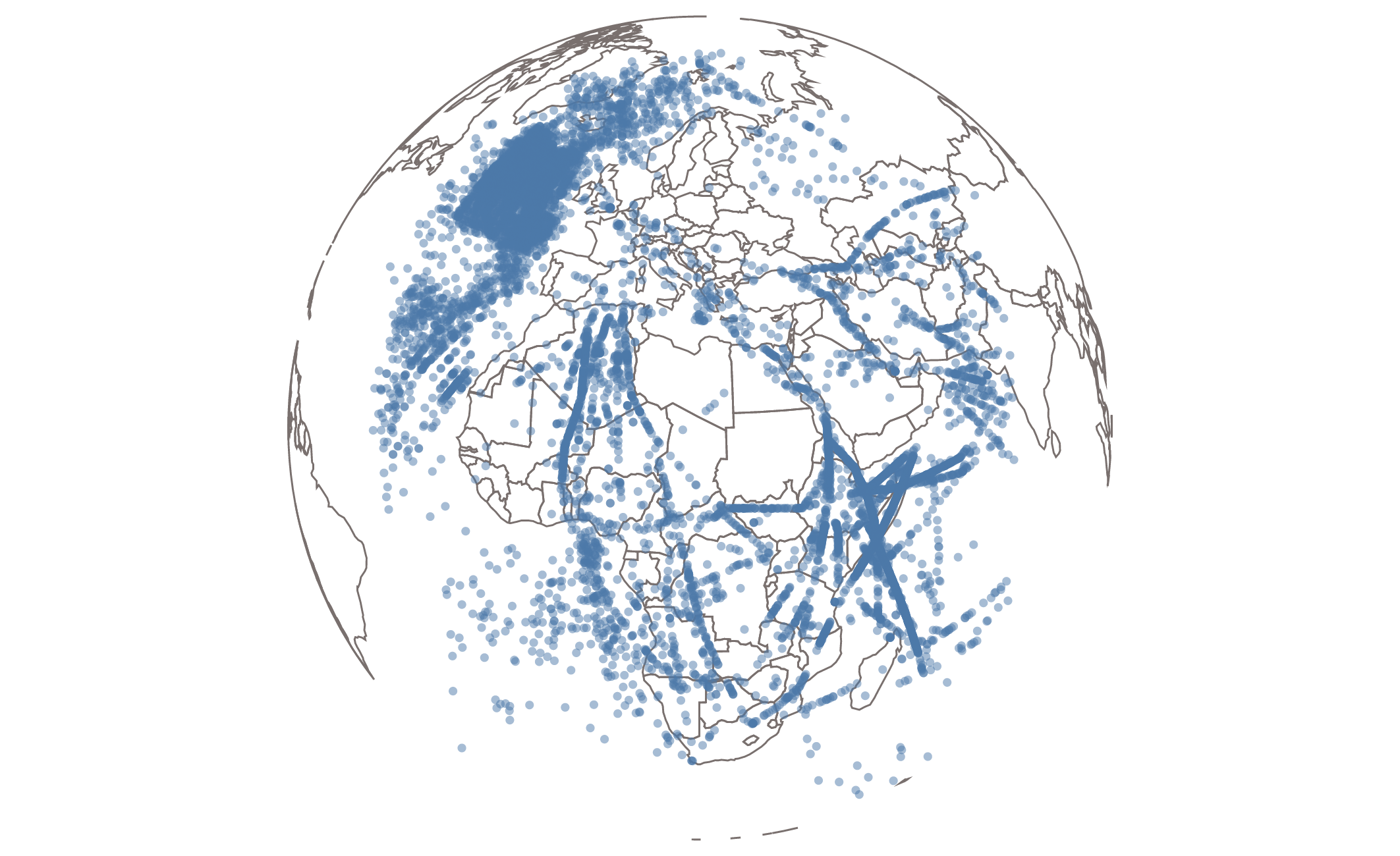}
    \caption{Illustration of the geographic origin of the received individual ADS-C messages}
    \label{fig:coverage}
\end{figure}

Figure \ref{fig:coverage} illustrates the coverage of the ADS-C reception in this dataset using all messages containing the position of the aircraft. We can see it spans most of the practical coverage of the Alphasat satellite in Figure \ref{fig:inmarsat}, ranging from Greenland to the Indian Ocean. Some main routes, as well as areas avoided by commercial flights during this time, such as Sudan and Lybia, are clearly visible.

\begin{figure}[!htbp]
    \centering
    \includegraphics[width=1\linewidth]{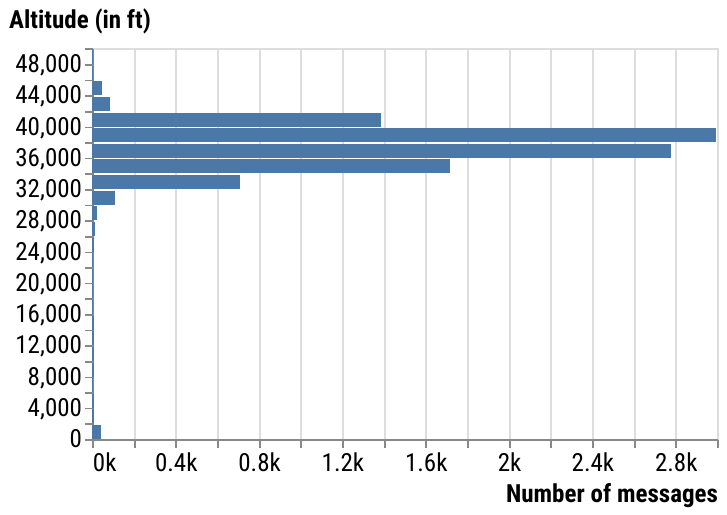}
    \caption{Histogram of the altitude reported in the received ADS-C messages}
    \label{fig:altitude}
\end{figure}

Finally, we analyze the altitude of the received aircraft positions. Figure \ref{fig:altitude} illustrates the clear focus of ADS-C on commercial jets during en-route flight phases. Aircraft are flying between 30,000 and 44,000 feet.

\section{Combining ADS-C Satellite Data with Ground-based ADS-B Data}
\label{sec:data_fusion}

The objective in this paper is to evaluate the potential of enhancing ADS-B trajectories with ADS-C data, particularly in areas where ADS-B coverage is limited. To construct complete flight trajectories, we combined satellite ADS-C data with historical data gathered by the OpenSky Network. It is worth noting that the ADS-C data is currently not available directly from OpenSky's Trino API, but it can be downloaded directly from an online open data repository \cite{xapelli_2025_14659997}. The historical state vector data from OpenSky Network is available through the Trino API.

To combine both data sources, the following data fusion process is applied:

\begin{itemize}
    \item Position data from ADS-C dataset is parsed using a parser we created for this purpose\footnote{\url{https://github.com/open-aviation/adscparse}}.
    \item Based on the callsign information, we split the ADS-C data points into individual flights.
    \item For each ADS-C flight, we query the related ADS-B data with the same transponder code using OpenSky's Trino API.
    \item The combined data per flight is filtered and resampled. 
\end{itemize}

\begin{figure}[!htbp]
    \centering
    \includegraphics[trim=150 50 150 50, clip, width=\linewidth]{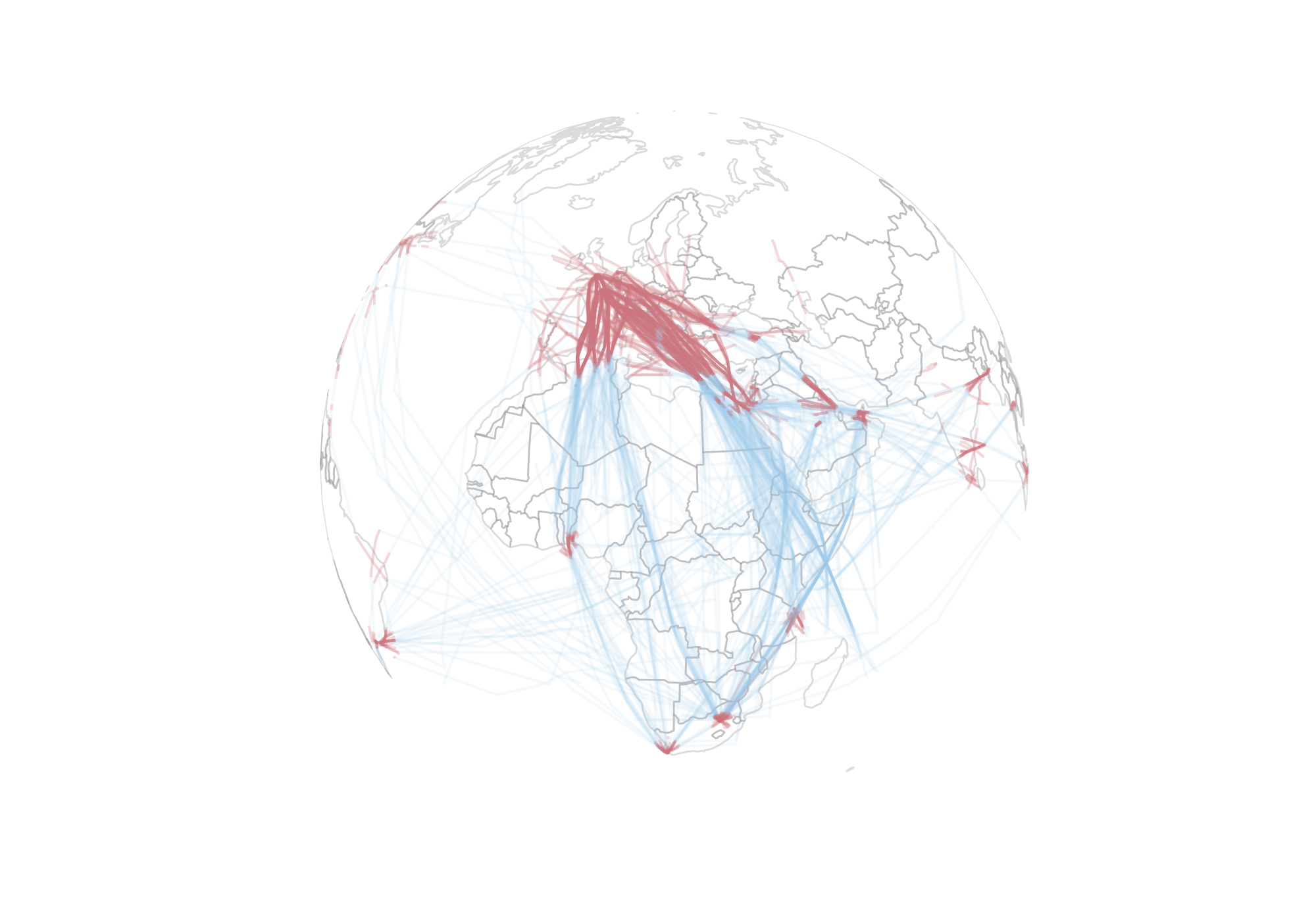}
    \caption{The globe shows a sample of 500 trajectories combining ADS-C (blue) and ADS-B (red) data}
    \label{fig:adsb_adsc_map}
\end{figure}

In Figure \ref{fig:adsb_adsc_map}, a random sample of 500 of these trajectories is illustrated. We can clearly see the ADS-B coverage in Europe and several major metropolitan centers in Africa and Asia, which leaves vast areas of Africa and the oceans uncovered and in need of trajectory estimation techniques such as a naive great circle distance measurement.

\section{Applications of the combined dataset}
\label{sec:applications}

\subsection{Better quantification of flight emissions}
In our previous work, we have conducted the estimation of fuel consumption and CO\textsubscript{2} emissions based on historical flights from OpenSky Network \cite{sun2022opensky}. However, due to the missing segment over the North Atlantic, we had to assume a great circle and interpolate the flight altitude based on only ADS-B flights. With the ADS-C data, we can now improve the accuracy of the fuel consumption and CO\textsubscript{2} emissions estimation, even with limited update rates of ADS-C over the oceanic regions.

\begin{figure}[!htbp]
    \centering
    \includegraphics[width=1\linewidth]{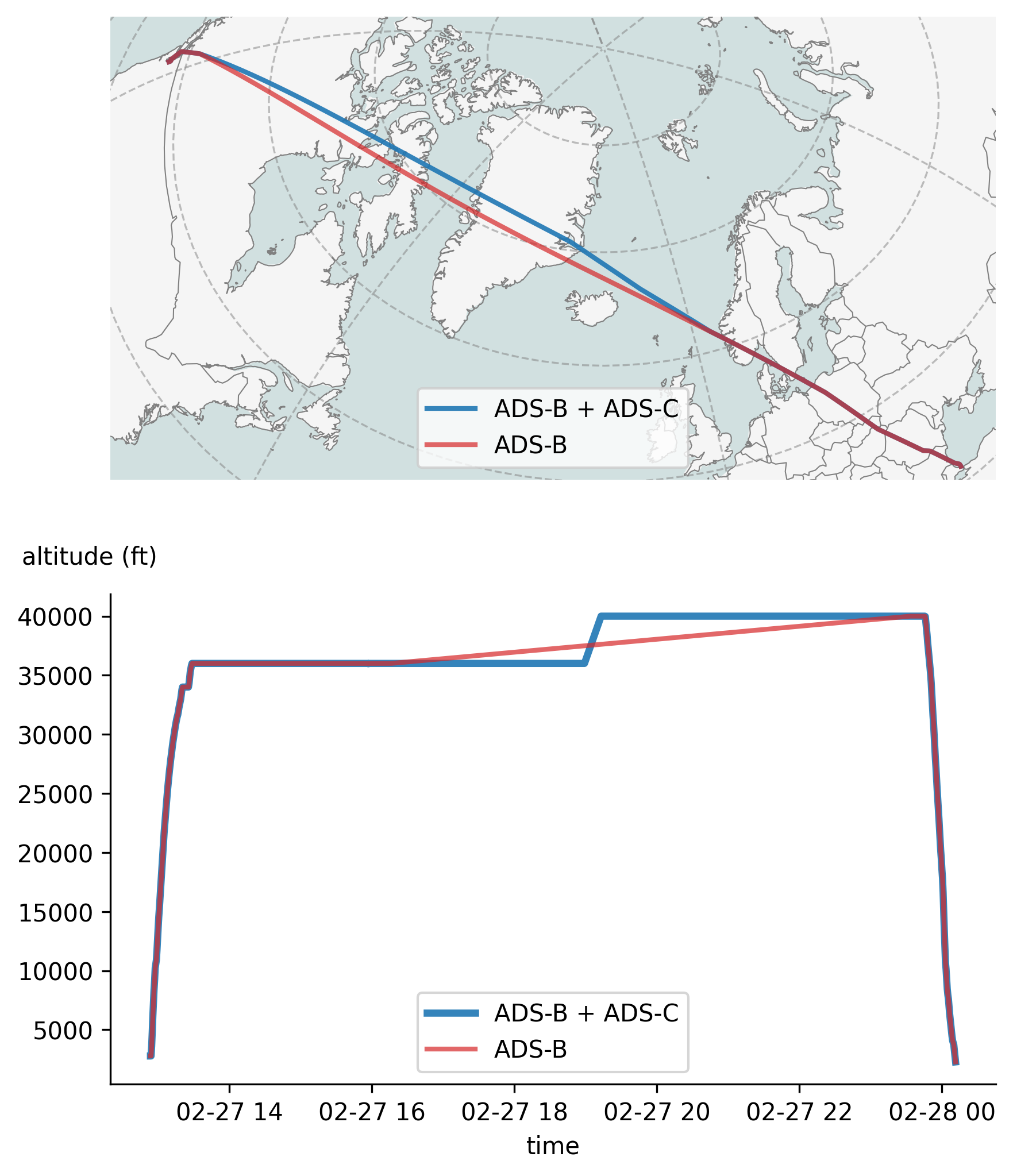}
    \caption{Comparison of ADS-B and combined ADS-B and ADS-C data for a single flight}
    \label{fig:example_comparison}
\end{figure}

\begin{figure}[!htbp]
    \centering
    \includegraphics[width=1\linewidth]{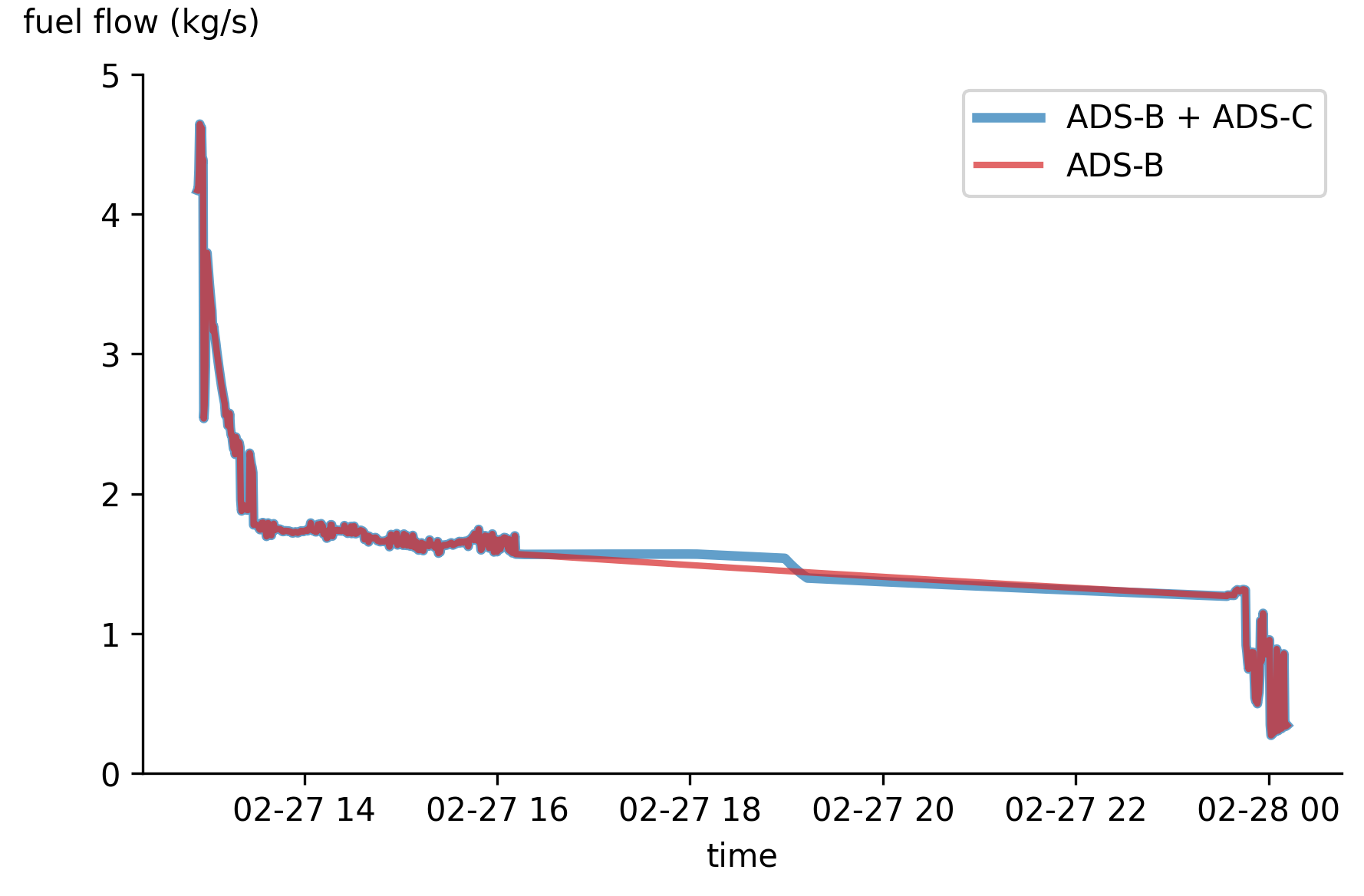}
    \caption{Comparison of fuel consumption estimates based on ADS-B and ADS-C data for a single flight}
    \label{fig:example_comparison_fuel}
\end{figure}

We illustrate the difference between ADS-B and combined ADS-B and ADS-C data for a single flight in Fig. \ref{fig:example_comparison}. We can see that the combined data provides a more accurate representation of the flight path. In Fig. \ref{fig:example_comparison_fuel}, we compare the fuel consumption estimates based on ADS-B and ADS-C data. During the cruise phase, the difference in fuel estimation can be significant, especially when the altitude of the flight is unknown and when the flight path does not follow the great circle flight path.

\subsection{Analysis of ADS-C flight in Africa}

\begin{figure}[!htbp]
    \centering
    \includegraphics[trim=150 50 150 50, clip, width=\linewidth]{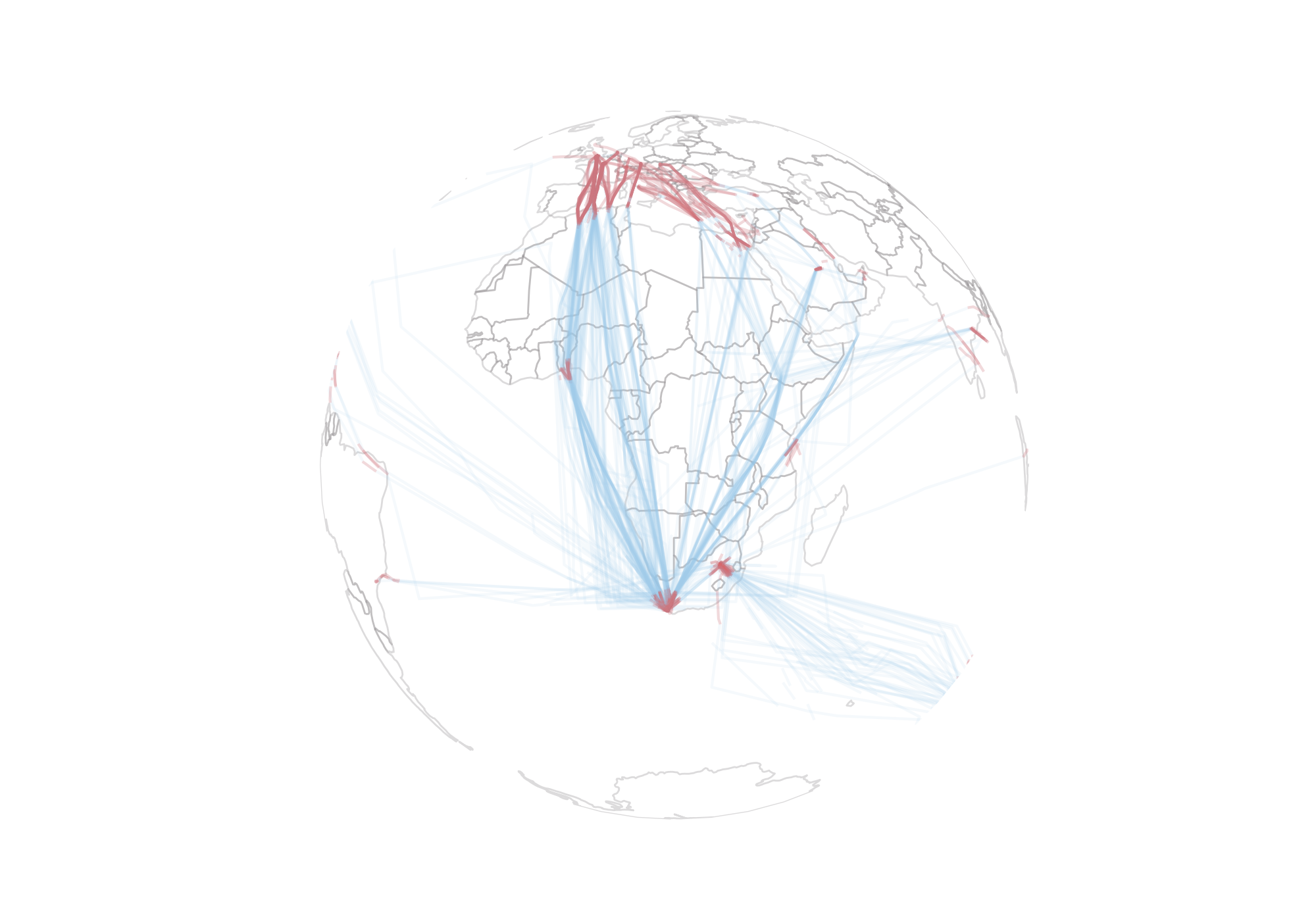}
    \caption{Illustration of improved coverage of trajectories from/to South Africa. ADS-C data is shown in blue, while ADS-B data is shown in red}
    \label{fig:sa-trajectory-sample}
\end{figure}

With the ADS-C data, we now have better coverage of flights in Africa, which are primarily long-haul flights with ADS-C capabilities. In  Fig. \ref{fig:sa-trajectory-sample}, we show another sample of trajectories from/to South Africa. It can be observed that the coverage of the trajectories is significantly improved with the addition of ADS-C data (shown in blue) as compared to the ADS-B data alone (shown in red).

\begin{figure}[!htbp]
    \centering
    \includegraphics[width=1\linewidth]{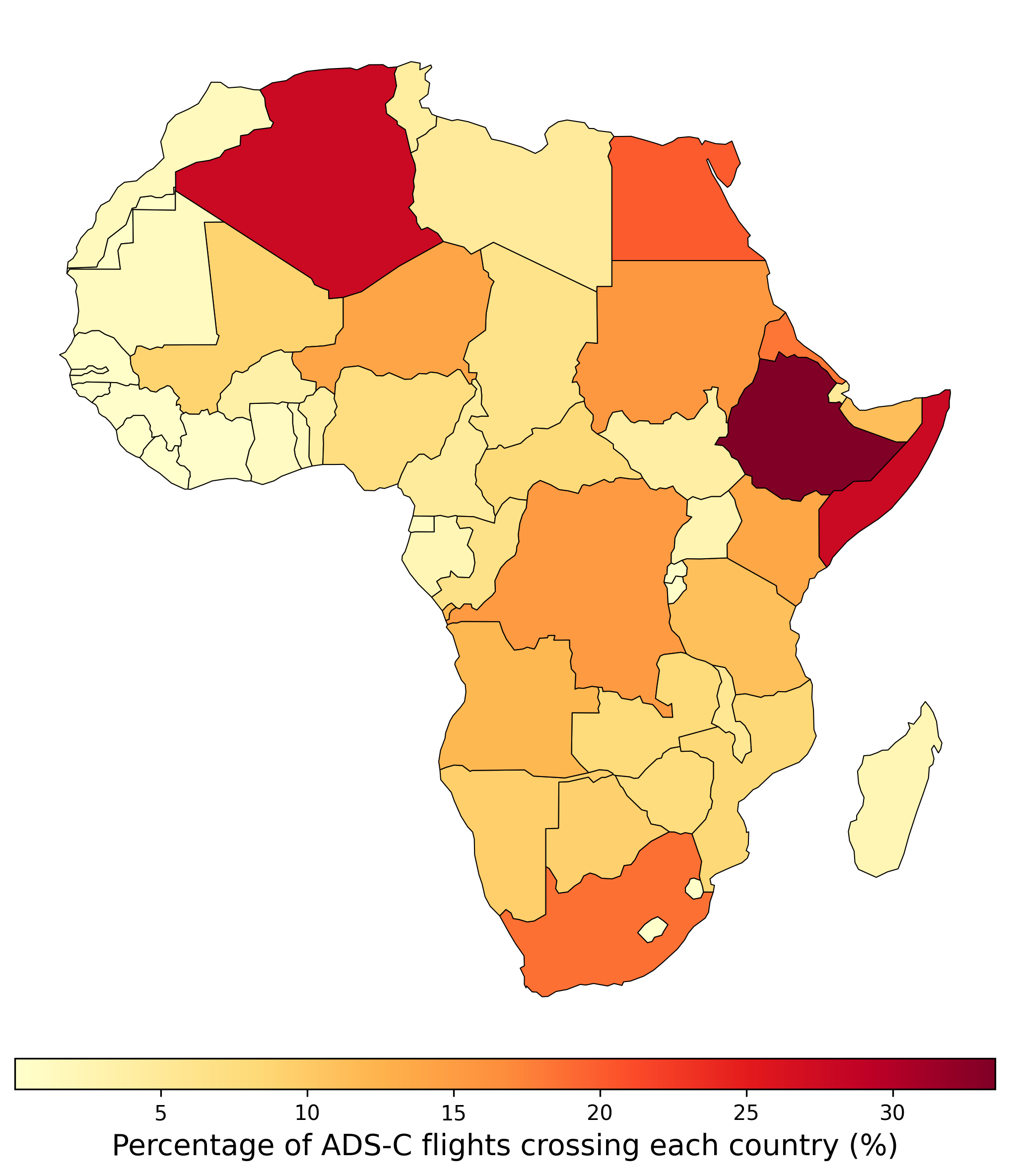}
    \caption{The percentage of the ADS-C flights crossing each country in Africa over 2024. Note that the total percentage of flights does not sum to 100\% as flights can cross multiple countries}
    \label{fig:africa_flights_percentage}
\end{figure}

We also extracted and analyzed the flights in Africa for the entire year. Table \ref{tab:top_countries} shows the top 15 African countries by the number of ADS-C flights. Ethiopia, Algeria, Somalia, Egypt, and South Africa are the top countries with the most cruise ADS-C flights. In Fig. \ref{fig:africa_flights_percentage}, we illustrate the percentage of the flights crossing each country in Africa. Note that the total percentage does not sum to 100\% as flights can cross multiple countries.

\begin{table}[!htbp]
\caption{Top 15 African Countries by Number of ADS-C Flights (percentages indicate the proportion of all flights that cross each country)}
\label{tab:top_countries}
\begin{tabular}{lcrr}
\toprule
Country & Country Code & ADS-C Flights & Percentage \\
\midrule
Ethiopia & ETH & 2569 & 33.50\% \\
Algeria & DZA & 2134 & 27.83\% \\
Somalia & SOM & 2128 & 27.75\% \\
Egypt & EGY & 1539 & 20.07\% \\
South Africa & ZAF & 1428 & 18.62\% \\
Eritrea & ERI & 1398 & 18.23\% \\
Sudan & SDN & 1194 & 15.57\% \\
Dem. Rep. Congo & COD & 1171 & 15.27\% \\
Niger & NER & 1072 & 13.98\% \\
Kenya & KEN & 1056 & 13.77\% \\
Angola & AGO & 916 & 11.94\% \\
Somaliland & -99 & 869 & 11.33\% \\
Tanzania & TZA & 845 & 11.02\% \\
Namibia & NAM & 724 & 9.44\% \\
Botswana & BWA & 704 & 9.18\% \\
\bottomrule
\end{tabular}
\end{table}

\section{Limitations and future work}
\label{sec:limitations}

Despite the advantages of the improved coverage with ADS-C data, several limitations should be acknowledged. The coverage of this dataset is primarily limited to the EMEA (Europe, the Middle East, and Africa) region, as it relies on data received by the Alphasat satellite, which does not provide global coverage. This geographic constraint means that flights in other regions, such as East Asia, Australia, and the Americas, are not well represented in our analysis.

Within Europe itself, we observed that aircraft rarely utilize ADS-C when flying over the continent. This fact is primarily because the dense network of ground-based ADS-B receivers provides comprehensive coverage, making satellite-based surveillance redundant in these areas. Consequently, our ADS-C dataset contains relatively few messages from aircraft during their continental European flight segments.

In Africa, while ADS-C is used more frequently due to the sparse ground-based surveillance infrastructure, the coverage is still restricted to a relatively small number of aircraft. Many African carriers do not have the necessary equipment to transmit ADS-C messages, resulting in significant gaps in the data across certain regions and routes, mainly for domestic flights within the continent. To improve the overall coverage of global south, deployment of more ground-based ADS-B receivers is still a desirable approach.

When examining transatlantic flights, we found that although ADS-C is used by many aircraft crossing the North Atlantic, the Alphasat satellite only covers approximately half of this oceanic area. This particular coverage creates a partial view of transatlantic traffic patterns, with a better coverage of eastern Atlantic routes than Western ones. Aircraft on polar routes or Western Atlantic tracks may be underrepresented in our dataset, potentially skewing analyses of overall traffic flows between Europe and North America.

To achieve full global coverage, we would need to establish a small network of ADS-C receivers covering the Pacific Ocean and the Americas. Current Inmarsat satellites undergo periodic orbital shifts in their geostationary positions. To maintain optimal receiving capability, satellite receivers should be able to track these movements and adjust their pointing direction accordingly. Low-cost solutions for such setups have already been proposed by the community\footnote{https://thebaldgeek.github.io/C-Band.html}. Extending of the ADS-C coverage will be the subject of ongoing and future work for the OpenSky Network.

\section{Conclusion}
\label{sec:conclusion}

In this paper, we have demonstrated the significant potential of combining ADS-B and ADS-C data to create more comprehensive flight trajectories. The OpenSky Network's integration of ADS-C data represents an important advancement in addressing the limitations of terrestrial ADS-B coverage, particularly over oceanic regions and remote areas such as Africa.

Our analysis of the ADS-C dataset reveals its valuable contribution to flight tracking, with over 720,000 messages from 2,606 unique aircraft. The predominance of basic reports and predicted route messages provides essential position data, while the coverage from various ATSUs, particularly Shanwick, Mogadishu, and Gander Oceanic, demonstrates the broad geographical reach of the dataset.

The combination of ADS-B and ADS-C data significantly improves trajectory reconstruction accuracy by up to 40\% for transoceanic flights and routes over Africa. Our findings address the central research question of how to overcome the inherent limitations of ground-based surveillance systems in remote areas. As demonstrated in our applications section, this enhanced dataset enables more accurate fuel consumption and emissions estimates by replacing simplistic great circle approximations with actual flight paths. Additionally, it provides unprecedented insight into flight patterns across the African continent, where traditional ADS-B coverage has been sparse, with over 33\% of flights crossing Ethiopia and nearly 28\% crossing Algeria and Somalia now properly represented in the data.

Despite these advantages, important limitations remain. The current dataset is primarily focused on the EMEA region due to Alphasat satellite coverage constraints. Many aircraft, particularly those operating domestic routes in Africa, may lack the ADS-C capability, and even the North Atlantic coverage is only partial. Future work should focus on expanding the receiver network to achieve more comprehensive global coverage, specifically by deploying low-cost satellite receiver solutions for Pacific and Americas coverage and improving data fusion algorithms to better integrate the different update rates between ADS-B and ADS-C.

The broader implications of this work extend beyond research to potential industry and policy impacts. More accurate flight tracking enables better emissions monitoring and reporting, supporting global climate initiatives.

The OpenSky Network's inclusion of ADS-C data represents another step forward in democratizing access to global flight data. While ADS-B and ADS-C differ in update rates, technical requirements, and aircraft equipage, they complement each other effectively to provide researchers with a more complete picture of global air traffic. By making this combined dataset openly available, we enable new research opportunities in air traffic management, environmental impact assessment, and aviation safety, particularly in previously underrepresented regions of the world.


\bibliography{reference}

\begin{thebibliography}{10}
\providecommand{\url}[1]{#1}
\csname url@samestyle\endcsname
\providecommand{\newblock}{\relax}
\providecommand{\bibinfo}[2]{#2}
\providecommand{\BIBentrySTDinterwordspacing}{\spaceskip=0pt\relax}
\providecommand{\BIBentryALTinterwordstretchfactor}{4}
\providecommand{\BIBentryALTinterwordspacing}{\spaceskip=\fontdimen2\font plus
\BIBentryALTinterwordstretchfactor\fontdimen3\font minus \fontdimen4\font\relax}
\providecommand{\BIBforeignlanguage}[2]{{%
\expandafter\ifx\csname l@#1\endcsname\relax
\typeout{** WARNING: IEEEtran.bst: No hyphenation pattern has been}%
\typeout{** loaded for the language `#1'. Using the pattern for}%
\typeout{** the default language instead.}%
\else
\language=\csname l@#1\endcsname
\fi
#2}}
\providecommand{\BIBdecl}{\relax}
\BIBdecl

\bibitem{schafer2019opensky}
M.~Schafer, X.~Olive, M.~Strohmeier, M.~Smith, I.~Martinovic, and V.~Lenders, ``{{OpenSky Report}} 2019: {{Analysing TCAS}} in the {{Real World}} using {{Big Data}},'' in \emph{Proceedings of the 38th {{IEEE}}/{{AIAA Digital Avionics Systems Conference}} ({{DASC}})}, San Diego, CA, Sep. 2019. doi: 10.1109/DASC43569.2019.9081686

\bibitem{olive2020opensky}
X.~Olive, A.~Tanner, M.~Strohmeier, M.~Schafer, M.~Feridun, A.~Tart, I.~Martinovic, and V.~Lenders, ``{{OpenSky Report}} 2020: {{Analysing}} in-flight emergencies using big data,'' in \emph{Proceedings of the 39th {{IEEE}}/{{AIAA Digital Avionics Systems Conference}} ({{DASC}})}, Sep. 2020. doi: 10.1109/DASC50938.2020.9256787

\bibitem{sun2021opensky}
J.~Sun, X.~Olive, M.~Strohmeier, M.~Schafer, I.~Martinovic, and V.~Lenders, ``{{OpenSky Report}} 2021: {{Insights}} on {{ADS-B Mandate}} and {{Fleet Deployment}} in {{Times}} of {{Crisis}},'' in \emph{Proceedings of the 40th {{IEEE}}/{{AIAA Digital Avionics Systems Conference}} ({{DASC}})}, San Antonio, TX, 2021. doi: 10.1109/DASC52595.2021.9594361

\bibitem{sun2022opensky}
J.~Sun, L.~Basora, X.~Olive, M.~Strohmeier, M.~Schafer, I.~Martinovic, and V.~Lenders, ``{{OpenSky Report}} 2022: {{Evaluating Aviation Emissions Using Crowdsourced Open Flight Data}},'' in \emph{Proceedings of the 41th {{IEEE}}/{{AIAA Digital Avionics Systems Conference}} ({{DASC}})}, Portsmouth, VA, Sep. 2022. doi: 10.1109/DASC55683.2022.9925852

\bibitem{sun2024opensky}
J.~Sun, X.~Olive, E.~Roosenbrand, C.~Parzani, and M.~Strohmeier, ``{{OpenSky Report}} 2024: {{Analysis}} of {{Global Flight Contrail Formation}} and {{Mitigation Potential}},'' in \emph{Proceedings of the 43th {{IEEE}}/{{AIAA Digital Avionics Systems Conference}} ({{DASC}})}, San Diego, CA, Oct. 2024.

\bibitem{olive2023opensky}
X.~Olive, J.~Sun, M.~Strohmeier, and G.~Tresoldi, ``{{OpenSky Report}} 2023: {{Low Altitude Traffic Awareness}} for {{Light Aircraft}} with {{FLARM}},'' in \emph{Proceedings of the 42th {{IEEE}}/{{AIAA Digital Avionics Systems Conference}} ({{DASC}})}, Barcelona, Spain, Oct. 2023. doi: 10.1109/DASC58513.2023.10311202

\bibitem{sun20232evaluating}
J.~Sun, A.~Tassanbi, P.~Obojski, and P.~Plantholt, ``Evaluating transatlantic flight emissions and inefficiencies using space-based {ADS-B} data,'' in \emph{Proc. of the 13th SESAR Innovation Days, Sevilla, Spain}, Dec. 2023. doi: 10.61009/SID.2023.1.41

\bibitem{xapelli2023first}
M.~Xapelli, T.~L{\"u}scher, G.~Tresoldi, M.~Strohmeier, and V.~Lenders, ``A first look at exploiting the automatic dependent surveillance-contract protocol for open aviation research,'' \emph{Journal of Open Aviation Science}, vol.~1, no.~2, 2023. doi: 10.59490/joas.2023.7229

\bibitem{strohmeier2017large}
M.~Strohmeier, ``Large-scale analysis of aircraft transponder data,'' \emph{IEEE Aerospace and Electronic Systems Magazine}, vol.~32, no.~1, pp. 42--44, 2017.

\bibitem{arinc-745}
{Airlines Electronic Engineering Committee}, \emph{{ARINC} characteristic 745-2: Automatic Dependent Surveillance ({ADS})}, 2nd~ed., 1993.

\bibitem{do-258a}
{Radio Technical Commission for Aeronautics}, \emph{{DO-258A: Interoperability Requirements for ATS Applications Using ARINC 622 Data Communications (FANS 1/A Interop Standard)}}, 2005.

\bibitem{xapelli_2025_14659997}
\BIBentryALTinterwordspacing
M.~Xapelli, M.~Strohmeier, and T.~Lüscher, ``{ADS-C} air traffic data collected by the {OpenSky Network},'' Jan. 2025. [Online]. Available: \url{https://doi.org/10.5281/zenodo.14659997}
\BIBentrySTDinterwordspacing

\end{thebibliography}
\bibliographystyle{ieeedoi}

\end{document}